\documentclass[a4paper]{article}
\usepackage{amsmath,amssymb}
\usepackage{setspace}
\usepackage{multirow}
\usepackage{multicol}
\usepackage{bm}
\usepackage{natbib}
\usepackage{color}
\usepackage[dvipsnames]{xcolor}
\usepackage{hyperref}
\definecolor{darkblue}{rgb}{0, 0, 0.5}
\hypersetup{
     colorlinks = true,
     citecolor = Maroon,
     urlcolor = darkblue,
     linkcolor = darkblue
}
\usepackage[a4paper,margin=2cm]{geometry}

\newcommand{\sbf}{\mathbf{s}}
\newcommand{\deltabf}{\boldsymbol{\delta}}
\newcommand{\R}{\mathbb{R}}
\newcommand{\M}{\mathcal{M}}

\begin{document}
\onehalfspacing

\title{Efficient counterfactual estimation in \\semiparametric discrete choice models:\\a note on \citet*{chiong2017counterfactual}}
\author{Grigory Franguridi\thanks{
		Department of Economics, University of Southern California.
		Email: \href{franguri@usc.edu}{franguri@usc.edu}
	}}

\date{\today}

\maketitle
\begin{abstract}
\small
I suggest an enhancement of the procedure of  \citet*{chiong2017counterfactual} for calculating bounds on counterfactual demand in semiparametric discrete choice models. Their algorithm relies on a system of inequalities indexed by \emph{cycles} of a large number $M$ of observed markets, and hence seems to require computationally infeasible enumeration of all such cycles. I show that such enumeration is unnecessary because solving the ``fully efficient'' inequality system exploiting cycles of all possible lengths $K=1,\dots,M$ can be reduced to finding the length of the shortest path between every pair of vertices in a complete bidirected weighted graph on $M$ vertices. The latter problem can be solved using the Floyd--Warshall algorithm with computational complexity $O\left(M^3\right)$, which takes only seconds to run even for thousands of markets. Monte Carlo simulations illustrate the efficiency gain from using cycles of all lengths, which turns out to be positive, but small.
\end{abstract}

\section{The Algorithm}

The family of inequalities for the linear program in \citet{chiong2017counterfactual} takes the form
\begin{align}
    \left(\deltabf^{l_1}-\deltabf^{M+1}\right)'\sbf^{M+1} &\leq -1'(D_1 \circ S_1) 1 = \sum_{k=1}^{K-1} \left(\deltabf^{l_k}-\deltabf^{l_{k+1}}\right)'\sbf^{l_k} \notag \\ &= \sum_{k=1}^{K-2} \left(\deltabf^{l_k}-\deltabf^{l_{k+1}}\right)'\sbf^{l_k} + \left(\deltabf^{l_{K-1}}-\deltabf^{M+1}\right)'\sbf^{l_{K-1}} \label{E:main_ineq}
\end{align}
for every cycle $l_1,\dots,l_{K},l_1$ from $\mathcal{M}= \{1,\dots,M\}$ with $l_K=M+1$. Here $\deltabf^m \in \R^J$ and $\sbf^m \in \Delta_J$ are the mean utility vector and the market share vector for market $m\in \mathcal{M}$, $\deltabf^{M+1}$ is the mean utility vector for the counterfactual market $M+1$ and $\sbf^{M+1}$ is the choice variable -- the counterfactual market share to be bounded. Denote the inequality \eqref{E:main_ineq} by $\text{ineq}(l_1,\dots,l_{K-1})$.

The number of possible cycles in a complete graph is exponential in $M$ and is of order $10^{20}$ already for $M=22$\footnote{See sequence A119913 in the Online Encyclopedia of Integer Sequences}. Therefore, simple enumeration is computationally infeasible and \citet{chiong2017counterfactual} resort to using a (tiny) subset of inequalities \eqref{E:main_ineq} corresponding to cycles of length $K=2$. I will show that enumeration is not necessary for this problem and that the full system of inequalities is easy to construct using graph optimization.

Define $G$ to be a complete bidirected weighted graph on vertices $\mathcal{M}$ with weight $w_{ij} = \left(\deltabf^{i}-\deltabf^{j}\right)'\sbf^{i}$ assigned to edge $(i,j) \in \mathcal{M}\times \mathcal{M}$. Since the mean utilities and shares for markets $\mathcal{M}$ are identified via cyclic monotonicity, it holds that
\begin{align*}
    \sum_{k=1}^{K} \left(\deltabf^{l_k}-\deltabf^{l_{k+1}}\right)'\sbf^{l_k} \geq 0
\end{align*}
for every path $l_1,\dots,l_{K},l_{K+1}=l_1$ from $\mathcal{M}$, and hence $G$ has no negative cycles. 

Note that the left-hand side of the inequality \eqref{E:main_ineq} depends on $\deltabf^{l_1}$ and so the inequalities for different values of $l_1$ are generally non-parallel. Hence, for each $l_1$, we can minimize the right-hand side over all possible paths $l_1,\dots,l_{K-1}$ starting from $l_1$ to get the sharpest inequality.

This minimization problem resembles the problem of finding the shortest path from vertex $l_1$ in the graph $G$, but is not equivalent to it because of the presence of the second term in \eqref{E:main_ineq}, corresponding to the last edge in the path.

I suggest iterating through $M(M-1)$ ordered pairs $(l_1,l_{K-1})$ of different vertices from $\M$ and, for each such pair, finding the shortest path that connects $l_1$ and $l_{K-1}$ in graph $G$. This can be accomplished using the Floyd--Warshall algorithm, which has polynomial computational complexity $O\left(M^3\right)$ for graphs with no negative cycles such as $G$. The length of the shortest path between $l_1$ and $l_{K-1}$ corresponds to the sharpest of the family of inequalities 
\begin{align*}
    \mathcal{F}_{l_1,l_{K-1}} = \left\{\text{ineq}(l_1,\dots,l_{K-1}): \,\,\, l_2,\dots,l_{K-2} \in \mathcal{M} \text{ s.t. } l_1,\dots,l_{K-1} \text{ is a path from } l_1 \text{ to } l_{K-1}\right\},
\end{align*}
from which we can then pick the sharpest one over $l_{K-1}$.

I suggest the following \textbf{algorithmic implementation}:
\begin{enumerate}
    \item Run the Floyd--Warshall algorithm on graph $G$, obtaining a matrix $D_G$ of shortest path lengths between every pair of vertices. Set the diagonal elements $(D_G)_{ii}=0$ for all $i\in\mathcal{M}$.
    \item For every $l_1\in \mathcal{M}$,
    \begin{itemize}
        \item for every $l\in \mathcal{M}$, set $\text{rhs}(l_1,l) = (D_G)_{l_1,l} + \left(\deltabf^{l}-\deltabf^{M+1}\right)'\sbf^{l}$,
        \item set $l^*=\arg \min_{l\in\mathcal{M}}\text{rhs}(l_1,l)$,
        \item save the inequality $\text{ineq}_{l_1}^*$ with the right hand side $\text{rhs}(l_1,l^*)$.
    \end{itemize}
    \item The resulting system of $M$ inequalities $\text{ineq}_{1}^*,\dots,\text{ineq}_{M}^*$ is equivalent to the system \eqref{E:main_ineq} with all cycles exhausted.
\end{enumerate}

\section{Monte Carlo Simulation}

I extend the Monte Carlo exercise of \citet{chiong2017counterfactual} and compare performance of their original procedure using only 2-cycles (denoted CHS) with the proposed procedure exhausting all possible cycles (denoted All cyc.).

I use the following two model specifications: one is multinomial logit as in \citet{chiong2017counterfactual} and the other is multinomial probit with the same DGP parameters as the logit and with the covariance matrix of the error term
\[
\Sigma=
\begin{pmatrix}
1 & -0.7 & 0.3\\
-0.7 & 1 & 0.3\\
0.3 & 0.3 & 1
\end{pmatrix}.
\]
A counterfactual of interest is the increase of price of good $j$ by 1\% (denoted $p_j\uparrow$), for $j=1,2,3.$ As can be seen from Tables \ref{tab:logit} and \ref{tab:probit}, the fully efficient procedure always dominates the CHS procedure both in terms of widths of the bounds and their standard errors. Although the gain is minimal and is unlikely to be relevant in practice, I suggest using the fully efficient procedure since it bears no extra computational costs (average computational time is less than 20 seconds for $M=1000$ and is less than 3 seconds for $M\in \{200,500\}$).

\begin{table}[ht]
\centering
\begin{tabular}{l c c c c | c c c | c c c}
& & \multicolumn{3}{c}{$M=200$} & \multicolumn{3}{c}{$M=500$} & \multicolumn{3}{c}{$M=1000$}\\
    \cline{3-11}
& & $s_1$ & $s_2$ & $s_3$ & $s_1$ & $s_2$ & $s_3$ & $s_1$ & $s_2$ & $s_3$ \\
\hline
\multirow{4}{*}{$p_1\uparrow$} & \multirow{2}{*}{CHS} & 0.0478&	0.0835& 0.0932 & 0.0074&	0.0373&	0.0367 & 0.0070& 0.0376&	0.0376\\
& &  (0.0136)&	(0.0267)	&(0.0290) & (0.0022)&	(0.0129)&	(0.0125) & (0.0030) &(0.0117)	&(0.0118) \\
\cline{2-11}
& \multirow{2}{*}{All cyc.} & 0.0454	& 0.0794	& 0.0878 & 0.0071 & 0.0351 & 0.0345 & 0.0066 &	0.0355 &	0.0355 \\
& & (0.0126) & (0.0243) &	(0.0255) & (0.0021)	&(0.0112)	&(0.0108) & (0.0026) &	(0.0103) &	(0.0104)\\
\hline
\multirow{4}{*}{$p_2\uparrow$} & \multirow{2}{*}{CHS} & 0.0890&	0.0575&	0.0978 & 0.0131	& 0.0245 & 0.0289 & 0.0101 & 0.0255 &	0.0289 \\
& & 0.0286 & 0.0217 & 0.0306 & 0.0046 &	0.0118 & 0.0105 & 0.0031 & 0.0075 &	0.0082 \\
\cline{2-11}
& \multirow{2}{*}{All cyc.} & 0.0837	& 0.0551 &	0.0915 & 0.0123 &	0.0230 & 0.0270 & 0.0096 & 0.0247 &	0.0278\\
& & (0.0256)& (0.0197) & (0.0271) & (0.0040) & (0.0098) & (0.0090) & (0.0028) & (0.0069) &	(0.0075)\\
\hline
\multirow{4}{*}{$p_3\uparrow$} & \multirow{2}{*}{CHS} & 0.0844 & 0.0833 & 0.0626 & 0.0125 &	0.0314 & 0.0268 & 0.0097 & 0.0263 &	0.0225\\
& & (0.0306) &	(0.0279) & (0.0254) & (0.0041) & (0.0085) &	(0.0073) & (0.0032) & (0.0092) &	(0.0086)\\
\cline{2-11}
& \multirow{2}{*}{All cyc.} & 0.0799 &	0.0790 & 0.0597 & 0.0119 &	0.0299 & 0.0256 & 0.0092 &	0.0250 &	0.0214\\
& & (0.0274) & (0.0256) & (0.0226) & (0.0036) & (0.0077)& (0.0066) & (0.0029) & (0.0083) &	(0.0077)\\
\hline
\end{tabular}

    \caption{Widths of bounds on counterfactual market shares and their standard errors. Bounds always cover true (\textbf{logit}) counterfactuals. Number of simulations is 100.}
    \label{tab:logit}
\end{table}

\begin{table}[ht]
\centering
\begin{tabular}{l c c c c | c c c | c c c}
& & \multicolumn{3}{c}{$M=200$} & \multicolumn{3}{c}{$M=500$} & \multicolumn{3}{c}{$M=1000$}\\
    \cline{3-11}
& & $s_1$ & $s_2$ & $s_3$ & $s_1$ & $s_2$ & $s_3$ & $s_1$ & $s_2$ & $s_3$ \\
\hline
\multirow{4}{*}{$p_1\uparrow$}	&	\multirow{2}{*}{CHS}	&	0.0050	&	0.0051	&	0.0003	&	0.0167	&	0.0516	&	0.0523	&	0.0006	&	0.0060	&	0.0061	\\
	&		&	(0.0015)	&	(0.0015)	&	(0.0002)	&	(0.0056)	&	(0.0161)	&	(0.0156)	&	(0.0001)	&	(0.0018)	&	(0.0018)	\\
	\cline{2-11}
	&	\multirow{2}{*}{All cyc.}	&	0.0048	&	0.0049	&	0.0003	&	0.0160	&	0.0487	&	0.0496	&	0.0006	&	0.0057	&	0.0058	\\
	&		&	(0.0013)	&	(0.0013)	&	(0.0002)	&	(0.0052)	&	(0.0140)	&	(0.0139)	&	(0.0001)	&	(0.0016)	&	(0.0016)	\\
	\hline
\multirow{4}{*}{$p_2\uparrow$}	&	\multirow{2}{*}{CHS}	&	0.0036	&	0.0035	&	0.0003	&	0.0271	&	0.0333	&	0.0434	&	0.0012	&	0.0048	&	0.0051	\\
	&		&	(0.0017)	&	(0.0017)	&	(0.0002)	&	(0.0082)	&	(0.0104)	&	(0.0133)	&	(0.0004)	&	(0.0012)	&	(0.0012)	\\
	\cline{2-11}
	&	\multirow{2}{*}{All cyc.}	&	0.0035	&	0.0033	&	0.0003	&	0.0256	&	0.0318	&	0.0412	&	0.0011	&	0.0046	&	0.0049	\\
	&		&	(0.0016)	&	(0.0015)	&	(0.0002)	&	(0.0075)	&	(0.0095)	&	(0.0121)	&	(0.0003)	&	(0.0011)	&	(0.0011)	\\
	\hline
\multirow{4}{*}{$p_3\uparrow$}	&	\multirow{2}{*}{CHS}	&	0.0051	&	0.0051	&	0.0000	&	0.0265	&	0.0413	&	0.0332	&	0.0013	&	0.0063	&	0.0063	\\
	&		&	(0.0021)	&	(0.0021)	&	(0.0000)	&	(0.0095)	&	(0.0137)	&	(0.0128)	&	(0.0004)	&	(0.0017)	&	(0.0017)	\\
	\cline{2-11}
	&	\multirow{2}{*}{All cyc.}	&	0.0049	&	0.0049	&	0.0000	&	0.0248	&	0.0389	&	0.0315	&	0.0012	&	0.0059	&	0.0059	\\
	&		&	(0.0020)	&	(0.0020)	&	(0.0000)	&	(0.0084)	&	(0.0122)	&	(0.0112)	&	(0.0004)	&	(0.0015)	&	(0.0016)	\\
\hline
\end{tabular}

    \caption{Widths of bounds on counterfactual market shares and their standard errors. Bounds always cover true (\textbf{probit}) counterfactuals. Number of simulations is 100.}
    \label{tab:probit}
\end{table}

\newpage

\bibliographystyle{ecta}
\bibliography{references}

\end{document}